\documentclass[twocolumn]{aastex62}
\usepackage{amsmath,color,textcomp,url,graphicx}
\usepackage{microtype}

\newcommand{\target}{2MASS~J13243553+6358281}
\newcommand{\shorttarget}{2MASS~J1324+6358}
\newcommand{\kms}{\hbox{km\,s$^{-1}$}}
\newcommand{\mjup}{$M_{\mathrm{Jup}}$}
\newcommand{\rjup}{$R_{\mathrm{Jup}}$}

\newcommand{\masyr}{$\mathrm{mas}\,\mathrm{yr}^{-1}$}
\newcommand{\teff}{$T_{\rm eff}$}
\newcommand{\lbol}{$\log{L_*/L_\odot}$}

\usepackage{color}
\definecolor{myred}{RGB}{200,0,0}

\received{January 29, 2018}
\revised{January 31, 2018}
\accepted{February 1, 2018}

\submitjournal{ApJL}

\shorttitle{2MASS~J13243553+6358281}
\shortauthors{Gagn\'e et al.}

\begin{document}

\title{2MASS~J13243553+6358281 IS AN EARLY T-TYPE PLANETARY-MASS OBJECT IN THE AB~DORADUS MOVING GROUP}

\author[0000-0002-2592-9612]{Jonathan Gagn\'e}
\affiliation{Carnegie Institution of Washington DTM, 5241 Broad Branch Road NW, Washington, DC~20015, USA}
\affiliation{NASA Sagan Fellow}
\email{jgagne@carnegiescience.edu}
\author[0000-0003-0580-7244]{Katelyn N. Allers}
\affiliation{Department of Physics and Astronomy, Bucknell University, Lewisburg, PA 17837, USA}
\author[0000-0002-9807-5435]{Christopher A. Theissen}
\affiliation{Center for Astrophysics and Space Sciences, University of California, San Diego, 9500 Gilman Dr., Mail Code 0424, La Jolla, CA 92093,
USA}
\author[0000-0001-6251-0573]{Jacqueline K. Faherty}
\affiliation{Department of Astrophysics, American Museum of Natural History, Central Park West at 79th St., New York, NY 10024, USA}
\author[0000-0001-8170-7072]{Daniella Bardalez Gagliuffi}
\affiliation{Department of Astrophysics, American Museum of Natural History, Central Park West at 79th St., New York, NY 10024, USA}
\author[0000-0002-8786-8499]{\'Etienne Artigau}
\affil{Institute for Research on Exoplanets, Universit\'e de Montr\'eal, D\'epartement de Physique, C.P.~6128 Succ. Centre-ville, Montr\'eal, QC H3C~3J7, Canada}

\begin{abstract}

We present new radial velocity and trigonometric parallax measurements indicating that the unusually red and photometrically variable T2 dwarf 2MASS~J13243553+6358281 is a member of the young ($\sim$\,150\,Myr) AB~Doradus moving group based on its space velocity. We estimate its model-dependent mass in the range 11--12\,\mjup\ at the age of \replaced{AB~Doradus}{the AB~Doradus moving group}, and its trigonometric parallax distance of $12.7 \pm 1.5$\,pc makes it one of the nearest known isolated planetary-mass objects. The \replaced{peculiar spectroscopic features of 2MASS~J13243553+6358281, such as its unusually red near-infrared continuum and inflated excess flux in the $K$ band, were}{unusually red continuum of 2MASS~J13243553+6358281 in the near-infrared was} previously suspected to be caused by an unresolved L+T brown dwarf binary, although it was never observed with high-spatial resolution imaging. This new evidence of youth suggests that a low surface gravity may be sufficient to explain \replaced{its peculiar features}{this peculiar feature}. Using the new parallax we find that its absolute $J$-band magnitude is $\sim$\,0.4\,mag fainter than equivalent-type field brown dwarfs, suggesting that the binary hypothesis is unlikely. The fundamental properties of 2MASS~J13243553+6358281 follow the spectral type sequence of other known high-likelihood members of \replaced{AB~Doradus}{the AB~Doradus moving group}. The effective temperature of 2MASS~J13243553+6358281 provides the first precise constraint on the L/T transition at a known young age, and indicates that it happens at a temperature of $\sim$\,1150\,K at $\sim$\,150\,Myr, compared to $\sim$\,1250\,K for field brown dwarfs.

\end{abstract}


\section{INTRODUCTION}\label{sec:intro}

The recent discovery of isolated planetary-mass objects (e.g., \citealp{2013ApJ...777L..20L,2015ApJ...808L..20G,2016ApJ...822L...1S,2016ApJ...821L..15K,2017ApJ...841L...1G}) and widely separated planetary-mass companions \citep{2005AA...438L..25C,2014ApJ...787....5N,2015ApJ...804...96G} in a range of ages provides a picture of how the atmospheres of planetary-mass objects evolve with time. The absence of a much brighter host star in their vicinity makes it possible to obtain high-resolution and high signal-to-noise spectra of their atmospheres with currently available facilities. These data are providing the foundations for our interpretation of the atmospheric features of directly imaged gas giant exoplanets such as 51~Eri~b \citep{2015Sci...350...64M}. Recent work has demonstrated overall similarities between such planetary-mass objects and giant planets at similar temperatures in their atmospheric features, but also significant differences caused mainly by their lower surface gravities. These include \deleted{weaker gravity-sensitive lines such as}\replaced{an inflated}{an excess flux in the} $K$ band caused by weakened collision-induced absorption of the H$_2$ molecule, \replaced{and}{compounded by} a redder \added{continuum} slope at near-infrared wavelengths \added{caused by thicker clouds} (e.g., see \citealp{2013ApJ...772...79A,2015ApJS..219...33G,2016ApJS..225...10F}). Recent work has also indicated tentative evidence that there may exist a correlation between low surface gravity and the presence of high-amplitude photometric variability \citep{2015ApJ...799..154M,2017AJ....154..138N,2017ApJ...841L...1G,2017arXiv171203746B}, but the small sample size still prevents a solid confirmation of this trend.

Because substellar objects cool down as they age \citep{2001RvMP...73..719B}, one of the main strategies to identify planetary-mass objects has been to focus on young associations with a well-known age, which provides a way to estimate masses based on observed temperatures and evolutionary models. The nearest young associations have a few members within 10\,pc of the Sun (e.g., \citealp{2017ApJ...841L...1G}), making them compelling laboratories to identify the lowest-mass objects in magnitude-limited surveys. However, their proximity means that their members are spread on large areas of the sky, making it hard to identify new members at a high confidence without measuring their full 6-dimensional kinematics. This includes the $XYZ$ Galactic coordinates and $UVW$ space velocity, which require measurements of the trigonometric parallax and radial velocity. Several methods were introduced to identify the most promising candidate members of young associations before measuring their radial velocity and distance (e.g., \citealp{2013ApJ...762...88M,2014ApJ...783..121G,2017AJ....153...95R,2017AJ....154...69S}), but all of them suffer from relatively low recovery rates and high rates of contamination for the nearest associations.

\begin{figure}
	\centering
	\includegraphics[width=0.485\textwidth]{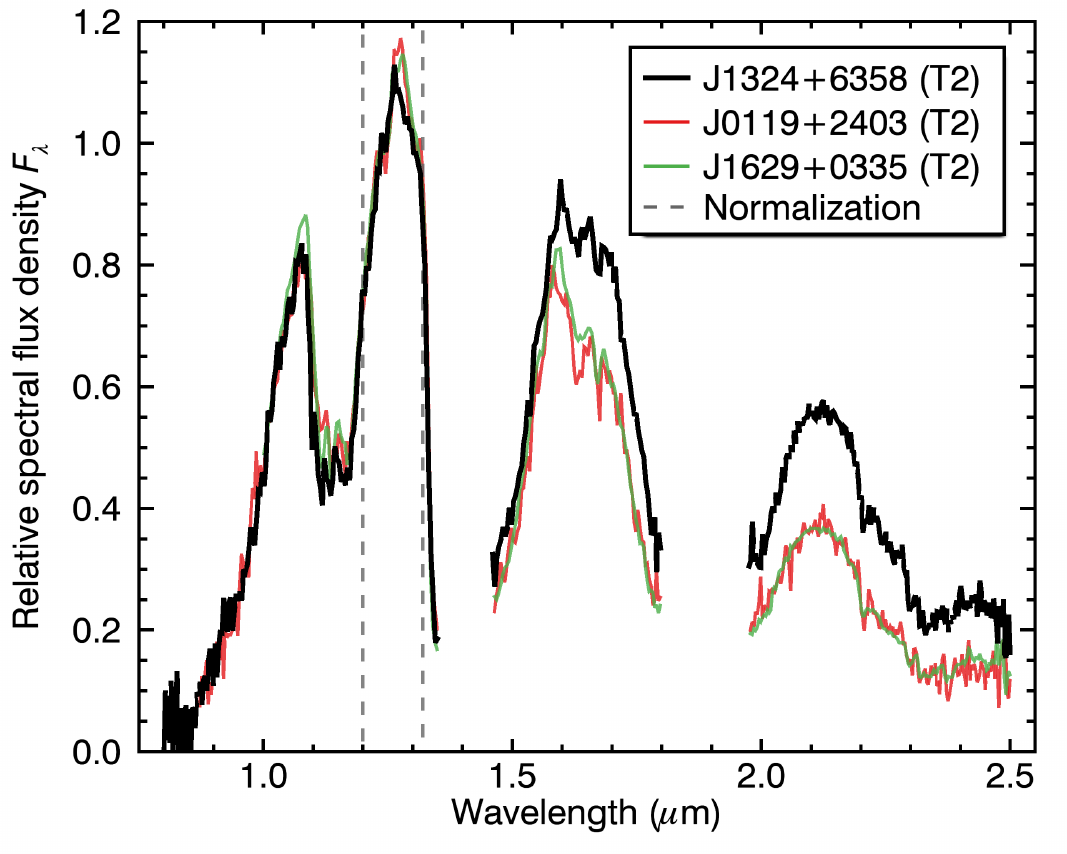}
	\caption{Relative spectral energy distribution of \shorttarget\ compared with two field T2 dwarfs. \shorttarget\ has a much redder slope\replaced{ and an inflated $K$ band}, which was previously identified as a sign of either a young age or an unresolved L9 + T2 binary.}
	\label{fig:spt}
\end{figure}

A new Bayesian tool, BANYAN~$\Sigma$ \citep{2018arXiv180109051G}, was recently built to identify members of young associations, with an updated method and new kinematic models for the associations based on the first data release of the Gaia mission \citep{2016AA...595A...4L} which provided 2 million parallaxes for nearby bright stars. This tool served as the base to start the BASS-Ultracool survey (J.~Gagn\'e et al., in preparation; see also \citealp{2015ApJ...808L..20G,2017ApJ...841L...1G}; and \citealt{Gagliuffi:2018uv}), an all-sky search for isolated planetary-mass objects in 2MASS \citep{2006AJ....131.1163S} and AllWISE \citep{2010AJ....140.1868W,2014ApJ...783..122K}, with a particular focus on objects in the T spectral class where clouds migrate to depths below their photosphere and methane becomes apparent in their spectra \citep{2006ApJ...637.1067B}. As part of this work, the known substellar object \target\ (\shorttarget\ hereafter) was identified as a candidate member of the $149_{-19}^{+51}$\,Myr-old \citep{2015MNRAS.454..593B} AB~Doradus moving group \replaced{\citep{2004ApJ...613L..65Z})}{(ABDMG; \citealt{2004ApJ...613L..65Z})}, based on its sky position and proper motion only.

Here, we report the first measurements of a trigonometric parallax and radial velocity based respectively on data from the \emph{WISE} survey and a new mid-resolution GNIRS spectrum, which indicate that \shorttarget\ has kinematics consistent with the locus of \replaced{AB~Doradus}{ABDMG} members. This is a strong indication that it is a young, low-gravity T dwarf in \replaced{AB~Doradus}{ABDMG}, and makes it unnecessary to invoke binarity to explain its peculiar spectroscopic features. We review relevant literature data on \shorttarget\ in Section~\ref{sec:lit}. In Section~\ref{sec:obs}, we detail the new GNIRS spectrum that was used to measure its radial velocity. The kinematics of \shorttarget\ are measured using these new data and \emph{WISE} astrometry in Section~\ref{sec:kinematics}, \replaced{and}{where} its membership to \replaced{AB~Doradus}{ABDMG} is \added{also }detailed\deleted{ in Section~\ref{sec:membership}}. The fundamental properties of \shorttarget\ are discussed in Section~\ref{sec:fundamental}, and this work is concluded in Section~\ref{sec:conclusion}.

\section{LITERATURE DATA}\label{sec:lit}

\shorttarget\ was identified as a substellar object by \cite{2007AJ....134.1162L}, and was noted for its \replaced{peculiar spectral features, notably its unusually red near-infrared slope and the inflated shape of its $K$ band a $\sim$\,2.15\,$\mu$m}{unusually red slope in the near-infrared} (see Figure~\ref{fig:spt}). It was hypothesized that the most likely causes of \replaced{these peculiar features}{this peculiar feature} were either that this object has a very young age (estimated at $<$\,300\,Myr) or that it consists of an unresolved binary with spectral types estimated at L9 + T2. \cite{2008ApJ...676.1281M} independently discovered this object and noted that its [4.5\,$\mu$m]$ - $[5.8\,$\mu$m] IRAC color is redder than any known T dwarf. \cite{2010ApJ...710.1142B} also posited that its peculiar spectral features may be due to binarity, but they noted that it was best fit by combining a T2 dwarf with 2MASS~J10430758+2225236 \citep{2007AJ....133..439C}, one of the reddest known objects classified as an optical L8 dwarf. No high-angular resolution imaging has been obtained yet for \shorttarget.

\cite{2014ApJ...783..121G} previously led an investigation of the potential young moving group membership of unusually red brown dwarfs with the BANYAN~II Bayesian classifier, and only obtained a 2.4\% probability that \shorttarget\ is a member of \replaced{AB~Doradus}{ABDMG}. However, they used near-infrared color magnitude diagrams that were extrapolated at spectral types $>$\,L7 because late-type young objects were unknown at the time, and they indicated that their results should be interpreted with care at these late spectral types. These color-magnitude diagrams were responsible for missing \shorttarget\ as a candidate member of \replaced{AB~Doradus}{ABDMG}, as ignoring them yields a 84\% membership with BANYAN~II using sky position and proper motion.

\section{OBSERVATIONS}\label{sec:obs}

An intermediate-resolution near-infrared spectrum was obtained for \shorttarget\ in good weather conditions with the GNIRS spectrograph \citep{2006SPIE.6269E..4CE} at the Gemini North telescope on UT 2018 January 5. Order 3 of the 110.5\,l/mm grating with a central wavelength setting of 2.3\,$\mu$m was used with the 0.05\,\arcsec/pix long camera and the 0\farcs15 slit, yielding a resolving power $\lambda/\Delta\lambda \sim $14,000 over 2.27--2.33\,$\mu$m. The target was nodded along the slit in a 6\arcsec-wide ABBA pattern for 13 exposures of 600\,s each, yielding a median signal-to-noise ratio per pixel of $\sim$\,40. Standard calibrations consisting of six 20\,s flat field exposures and two 80\,s Xe, Ar arc lamps were also obtained to correct pixel sensitivity and obtain an initial wavelength solution.

The data were reduced using a modified version of {\tt REDSPEC}, Keck Observatory's facility data reduction package for high-resolution, near-infrared spectra. The reduction included an initial wavelength calibration (using sky emission lines as well as Xe, Ar arc lamps), flat-fielding and bad-pixel masking. Each nod pair was subtracted to eliminate sky emission, and the resulting A-B images were spatially and spectrally rectified. Spectra were extracted from these A-B images using an aperture equal to 0.67 times the full width at half maximum of the seeing and subtracting residual sky lines using background windows without significant source flux. Uncertainties were calculated by propagating Poisson noise throughout the reduction process. The extracted spectra were combined using a robust weighted mean with the {\tt xcombspec} procedure from the {\tt Spextool} package \citep{2004PASP..116..362C}.

\section{KINEMATICS}\label{sec:kinematics}

\subsection{Radial Velocity}\label{sec:rv}

\begin{figure*}
	\centering
	\includegraphics[width=0.95\textwidth]{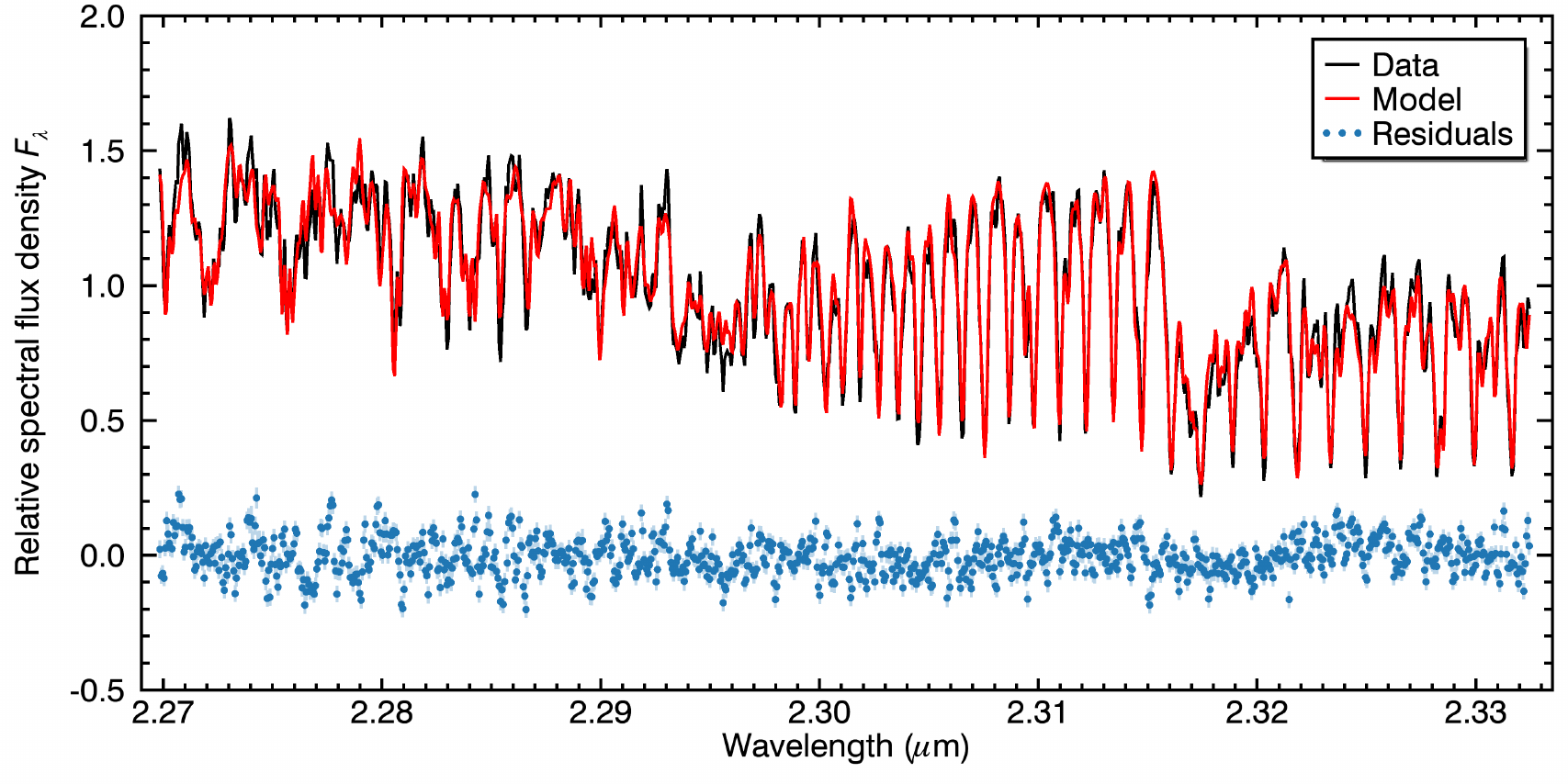}
	\caption{Best-fitting forward model (red) compared to the GNIRS spectrum of \shorttarget\ obtained here (black). Residuals are shown as blue dots. We obtain a reduced $\chi^2 = 2.3$.}
	\label{fig:rv}
\end{figure*}

A forward model for the GNIRS spectrum was built from radial velocity-shifted BT-Settl (CIFIST) atmosphere models \citep{2012RSPTA.370.2765A} convolved with a rotational broadening profile, including a second-degree polynomial wavelength solution. The {\tt atran} model for Mauna Kea \citep{1992nstc.rept.....L}\footnote{Calculated at a zenith angle of 48\textdegree\ with 3.1\,mm of water vapor, see \url{https://atran.sofia.usra.edu/cgi-bin/atran/atran.cgi}} was then included with optical depth as a free parameter to model telluric absorption, and the resulting spectrum was convolved with a Gaussian line spread function to represent instrumental effects. The best-fitting values for these 7 forward model parameters were explored with a DREAM-ZS Markov Chain Monte Carlo method \citep{terBraak:2008iw}, where values of $\log g$ and $T_{\rm eff}$ between grid points are explored by interpolating the nearest BT-Settl models on the grid. The corresponding Markov chains were used to build the probability density functions of the radial velocity and projected rotational velocity. A 4\% systematic uncertainty was included in the data before performing this analysis so that the median uncertainties of the data match the median absolute residuals of the best fit. This resulted in a radial velocity of $-23.7_{-0.2}^{+0.4}$\,\kms\ and a projected rotational velocity of $v\sin i = 11.5_{-0.8}^{+1.0}$\,\kms.\added{ The model $\log g$ and $T_{\rm eff}$ values obtained from this forward fitting are $5.30_{-0.01}^{+0.02}$\,dex and $1304.0_{-1.5}^{+1.7}$\,K respectively, however such measurements for substellar objects based on fitting atmosphere models are unreliable especially when obtained from a narrow wavelength range \citep{2015ApJS..219...33G,2015ApJ...810..158F,2016ApJ...819..133A}.} This method is described in detail in \cite{2016ApJ...819..133A}, where it is used to derive the radial velocity and projected rotational velocity of PSO~J318.5--22 from a GNIRS spectrum obtained in the same configuration. The resulting best-fitting model is presented in Figure~\ref{fig:rv}.

\begin{figure*}
	\centering
	\includegraphics[width=0.82\textwidth]{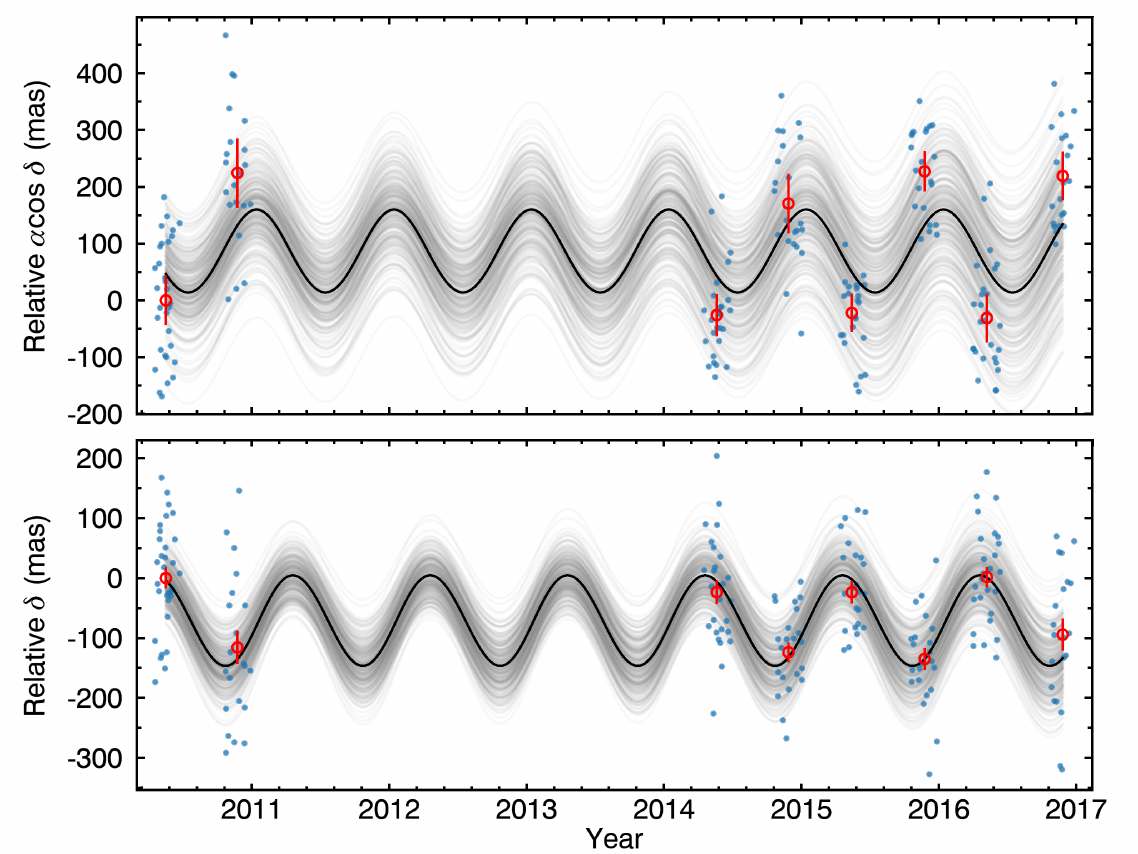}
	\caption{Relative astrometry of \shorttarget\ in right ascension and declination (red circles) with proper motion subtracted, compared to the best-fitting model (black line) and 300 random realizations selected in the Markov chains (gray lines). The blue dots are individual astrometric measurements used to build the 8 astrometric epochs with the method of \cite{2017arXiv171011127T}.}
	\label{fig:plx}
\end{figure*}

\begin{deluxetable}{lcc}
\tabletypesize{\small}
\tablecaption{Properties of \target \label{tab:properties}}
\tablehead{\colhead{Property} & \colhead{Value} & \colhead{Reference}}
\vspace{-0.2cm}\startdata
\sidehead{\textbf{Position and Kinematics}\vspace{-0.1cm}}
R.A. & 13:24:45.894 $ \pm $55\,mas & 1\\
Decl.  & +63:58:12.76 $ \pm $28\,mas & 1\\
Epoch (JD)  & 2455334.4 $ \pm $0.5\tablenotemark{a} & 1\\
$\mu_\alpha\cos\delta$ (\masyr) & $-379.7 \pm 7.7$ & 2\\
$\mu_\delta$ (\masyr) & $-66.2 \pm 7.0$ & 2\\
Radial velocity (\kms) & $-23.7_{-0.2}^{+0.4}$ & 1\\
Parallax (mas) & $78.7 \pm 9.0$ & 1\\
Trigonometric distance (pc) & $12.7 \pm 1.5$ & 1\\
$X$ (pc) & $-3.5 \pm 0.4$ & 1\\
$Y$ (pc) & $6.9 \pm 0.8$ & 1\\ 
$Z$ (pc) & $10.1 \pm 1.2$ & 1\\
$U$ (\kms) & $-9.5 \pm 1.9$ & 1\\
$V$ (\kms) & $-28.8 \pm 1.9$ & 1\\
$W$ (\kms) & $-13.6 \pm 0.7$ & 1\vspace{-0.2cm}\\
\sidehead{\textbf{Photometric Properties}\vspace{-0.1cm}}
$i_{\rm AB}$ (SDSS) & $22.73 \pm 0.28$ & 3\\
$z_{\rm AB}$ (SDSS) & $18.72 \pm 0.04$ & 3\\
$J$ (2MASS) & $15.60 \pm 0.07$ & 4\\
$H$ (2MASS) & $14.58 \pm 0.06$ & 4\\
$K$ (2MASS)& $14.06 \pm 0.06$ & 4\\
$W1$ (AllWISE) & $13.12 \pm 0.02$ & 5\\
$W2$ (AllWISE) & $12.29 \pm 0.02$ & 5\\
$W3$ (AllWISE) & $10.78 \pm 0.07$ & 5\\
\mbox{[3.6\,$\mu$m]} (IRAC) & $12.56 \pm 0.03$ & 6\\
\mbox{[4.5\,$\mu$m]} (IRAC) & $12.33 \pm 0.03$ & 6\\
\mbox{[5.8\,$\mu$m]} (IRAC) & $11.79 \pm 0.03$ & 6\\
\mbox{[8.0\,$\mu$m]} (IRAC) & $11.31 \pm 0.03$ & 6\vspace{-0.2cm}\\
\sidehead{\textbf{Fundamental Properties}\vspace{-0.1cm}}
Spectral type & T2\,$\pm 1$& 6,7\\
Age (Myr) & $149_{-19}^{+51}$ & 1,8\\
Mass (\mjup) & 11--12 & 1\\
Radius (\rjup) & $1.23 \pm 0.02$ & 1\\
\teff\ (K) & $1080 \pm 60$ & 1\\
$\log g$ & $4.29 \pm 0.02$ & 1\\
\lbol & $-4.72 \pm 0.10$ & 1\\
Rotation period (h) & $13 \pm 1$ & 9\\
$v\sin i$ (\kms) & $11.5_{-0.8}^{+1.0}$ & 1\\
$i$ (\textdegree) & $56_{-7}^{+11}$ & 1\vspace{0.1cm}\\
\enddata
\tablenotetext{a}{This corresponds to the first combined epoch in the WISE measurements, where the error bar is given by the standard deviation of the individual epochs that were combined.}
\tablerefs{(1)~This work, (2)~\citealt{2016ApJ...817..112S}, (3)~\citealt{2015ApJS..219...12A}, (4)~\citealt{2006AJ....131.1163S}, (5)~\citealt{2014ApJ...783..122K}, (6)~\citealt{2008ApJ...676.1281M}, (7)~\citealt{2007AJ....134.1162L}, (8)~\citealt{2015MNRAS.454..593B}, (9)~\citealt{2015ApJ...799..154M}}
\end{deluxetable}

\subsection{Trigonometric Parallax}\label{sec:tdist}

The trigonometric parallax of \shorttarget\ was measured with a slight adaptation of the method described by \cite{2017arXiv171011127T}. In summary, all individual astrometric epochs of \shorttarget\ in the $W1$ and $W2$ bands were collected from all phases of the \emph{WISE} mission\footnote{\emph{WISE} All-Sky, \emph{WISE} 3-Band Cryo, \emph{WISE} Post-Cryo and NEOWISE Reactivation.} (providing a total baseline of 6.5\,yr), and compared to a parallax and proper motion solution with the Markov Chain Monte Carlo Python package {\tt emcee} \citep{2013ascl.soft03002F}. The method of \cite{2017arXiv171011127T} has only 3 free parameters corresponding to the parallax and the two dimensions of proper motion, and compares the forward parallax model to all 8 astrometric epochs relative to the first measurement. To avoid propagating the information of the first measurement across all epochs, we included the average sky position of \shorttarget\ as two additional free parameters. This modification makes this analysis less vulnerable to outlier measurements, and yields more conservative error bars on all parameters. The resulting parallax measurement is $78.7 \pm 9.0$\,mas, with a proper motion of $\mu_\alpha\cos\delta = -378.3 \pm 10.4$\,\masyr\ and $\mu_\delta = -68.0 \pm 5.7$\,\masyr. The best-fitting astrometric solution is displayed in Figure~\ref{fig:plx}. As explained by \cite{2017arXiv171011127T}, this astrometric solution does not include a correction for the motion of background stars. \replaced{The resulting apparent to absolute correction is within}{They estimated a parallax correction of the order of 2\,mas, well within} the error bar for the parallax, but not for the proper motion, and we therefore adopt the proper motion measurement of \citeauthor{2016ApJ...817..112S} (\citeyear{2016ApJ...817..112S}; based on 2MASS and AllWISE with a 10.4\,yr baseline), which is reported in Table~\ref{tab:properties} with our parallax measurement.

\begin{figure*}
	\centering
	\includegraphics[width=0.95\textwidth]{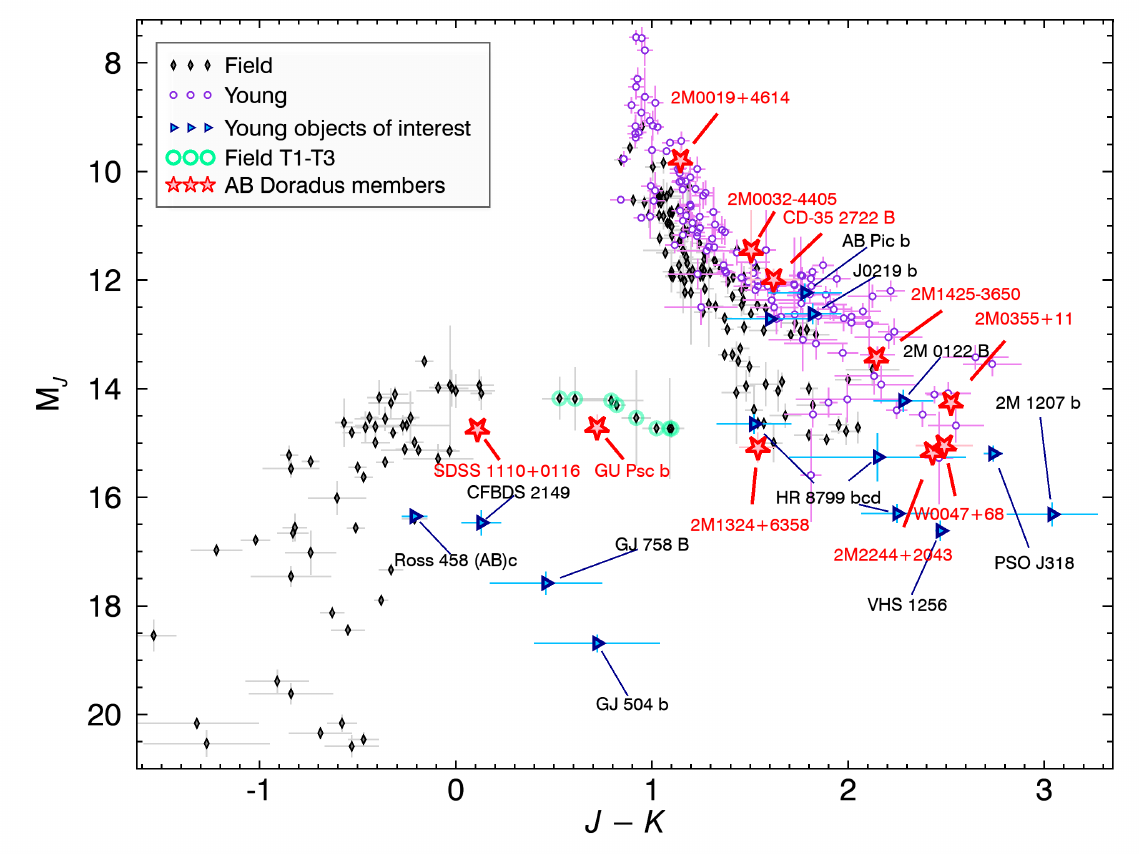}
	\caption{Color-magnitude diagram of confirmed or high-likelihood members of \replaced{AB~Doradus}{ABDMG} (as defined by \citealt{2016ApJS..225...10F}) with trigonometric parallax measurements (red stars) compared with field (black diamonds) and young brown dwarfs (open purple circles and right-pointing blue triangles). Field brown dwarfs within one spectral subtype of \shorttarget\ are marked with a green open circle. The \replaced{AB~Doradus}{ABDMG} members define a sequence redder than the field, with an L/T transition happening at a slightly fainter $J$-band magnitude. \shorttarget\ is both redder and fainter than field T1--T3 dwarfs in the near-infrared, similarly to young late-type L dwarfs. These data were compiled from \cite{2015ApJS..219...33G}, \cite{2016ApJS..225...10F}, \cite{2016ApJ...833...96L}, \cite{2018MNRAS.474.1041V} and other references therein.}
	\label{fig:cmd}
\end{figure*}

\subsection{Kinematic Membership to Nearby Young Associations}\label{sec:membership}

The BANYAN~$\Sigma$ algorithm was used to determine the Bayesian probability that \shorttarget\ is a member of young associations within 150\,pc of the Sun. The sky position, proper motion, radial velocity and trigonometric parallax distance of \shorttarget\ (see Table~\ref{tab:properties}) were used and yielded an \replaced{AB~Doradus}{ABDMG} membership probability of 98.0\%. The measured $UVW$ space velocity of \shorttarget\ (see Table~\ref{tab:properties}) is located within 1.1\,\kms\ of the locus of \replaced{AB~Doradus}{ABDMG} members compiled in \cite{2018arXiv180109051G}, within the velocity dispersion of the moving group (1.4\,\kms), corresponding to a 0.7$\sigma$ distance in the kinematics tri-dimensional space when accounting for the covariances in the distribution of existing \replaced{AB~Doradus}{ABDMG} members.\added{ The Galactic coordinates $XYZ$ of \shorttarget\ are located at $\sim$\,22.7\,pc from the core of ABDMG members, corresponding to a 1.1\,$\sigma$ distance from the core of the BANYAN~$\Sigma$ model. ABDMG has a relatively large characteristic size of $\sim$\,19\,pc in $XYZ$ space \citep{2018arXiv180109051G}.}

\section{FUNDAMENTAL PROPERTIES}\label{sec:fundamental}

\subsection{Color-Magnitude Diagram}

The trigonometric parallax measured in Section~\ref{sec:tdist} allows us to place \shorttarget\ in a color-magnitude diagram and compare it with the sequence of field brown dwarfs (Figure~\ref{fig:cmd}). \shorttarget\ falls in sequence with other young members of \replaced{AB~Doradus}{ABDMG}, redder in $J - K$ than field brown dwarfs, and where the L/T transition takes place at a slightly fainter $J$-band magnitude. This L/T transition magnitude dependence on age was hinted by the properties of the HR~8799~bcd planets \citep{2008Sci...322.1348M,2012ApJ...754..135M} and those of the \replaced{AB~Doradus}{ABDMG} members SDSSp~J111010.01+011613.1 and GU~Psc~b (\citealp{2015ApJ...808L..20G,2014ApJ...787....5N}; see also the discussion of \citealt{2016ApJ...833...96L}), but \shorttarget\ provides yet stronger evidence for this effect.

\subsection{Multiplicity}

The \deleted{slightly }fainter $J$-band magnitude of \shorttarget\ compared to the field substellar sequence goes against the spectral binary hypothesis, unless there is a large $J$-band contrast ratio between the two components, which would be inconsistent with the best-fitting L9 + T2 components identified by \cite{2007AJ....134.1162L} and \cite{2010ApJ...710.1142B}.\added{ If we assume that \shorttarget\ is an equal-mass binary with a conservative mass estimate in the range 5--60 $M_{\rm Jup}$ based on L9 + T2 spectral types at any age, our inability to resolve doubled spectral lines with the GNIRS spectrum corresponds to a weak lower limit in physical separation in the range ~0.02--0.22\,AU, or an angular separation of $\sim$\,1.6--17.3\,mas at the measured trigonometric parallax distance.} \added{The absence of elongation in the point-spread function of the GNIRS acquisition images\footnote{Available at the Gemini Science Archive at \url{https://archive.gemini.edu}} puts an upper limit on the separation of an equal-luminosity binary at $\sim$\,0\farcs2, or $\sim$\,2.5\,AU. }The fact that the kinematics of \shorttarget\ are consistent with a membership in \replaced{AB~Doradus}{ABDMG} is sufficient to explain its peculiar spectroscopic features, as described in Section~\ref{sec:lit}.\added{ As a consequence of these considerations, we consider the binary hypothesis as highly unlikely.} This is also consistent with the general picture that high variability in substellar objects seems to be correlated with low surface gravity and thus youth \citep{2015ApJ...799..154M,2017ApJ...841L...1G}. The case of \shorttarget\ would also be reminiscent of 2MASS~J21392676+0220226, which was first identified as a candidate L8.5 + T4.5 spectral binary by \cite{2010ApJ...710.1142B}. \cite{2012ApJ...750..105R} later ruled out the presence of a companion at separations above $\sim$\,1.6\,AU for a $J$-band constrast of 3\,mag or less using \emph{HST}/NICMOS imaging, and discovered it to be highly variable with a 26\% peak-to-peak $J$-band amplitude.  This suggested that the high variability of brown dwarfs may be due to evolving and rotating patterns of patchy clouds, presenting us a combination of more than a single atmospheric layer at different effective temperatures. We can therefore suspect that highly variable brown dwarfs will cause more false positives in the detection of binaries with the spectral fitting method.

\subsection{Spectral Energy Distribution}

The trigonometric parallax measured here was combined with the 0.8--2.5\,$\mu$m low-resolution near-infrared spectrum of \cite{2007AJ....134.1162L} and all photometric data listed in Table~\ref{tab:properties} (i.e. SDSS, 2MASS, AllWISE and IRAC) to build the spectral energy distribution of \shorttarget. The method described in \cite{2015ApJ...810..158F} and \cite{2016ApJS..225...10F} was then used to directly measure its bolometric luminosity, estimate a radius and a mass based on the \cite{2008ApJ...689.1327S} evolutionary models at the age of \replaced{AB~Doradus}{ABDMG}, and translate the radius and bolometric luminosity to a semi-empirical effective temperature using the Stefan-Boltzmann law. The resulting fundamental properties of \shorttarget\ are listed in Table~\ref{tab:properties}. \shorttarget\ falls in sequence with other high-likelihood \replaced{AB~Doradus}{ABDMG} members (see Figure~\ref{fig:cmd}) in spectral type, mass, temperature and \lbol\ measured with the same method \citep{2015ApJ...808L..20G,2016ApJS..225...10F}, although its mass estimate overlaps significantly with that of the L6--L8\,$\gamma$-type bona fide member 2MASS~J22443167+2043433 \citep{2018MNRAS.474.1041V}. The temperatures of 2MASS~J22443167+2043433 ($1180 \pm 10$\,K; \citealt{2016ApJS..225...10F}) and \shorttarget\ ($1080 \pm 60$\,K) provide an indication that the transition from the L to T spectral type happens at $\sim$\,1150\,K at the age of \replaced{AB~Doradus}{ABDMG}, a temperature slightly cooler than for field-aged brown dwarfs ($\sim$\,1250\,K; \citealt{2015ApJ...810..158F}). This is consistent with previous observations in the literature (e.g., see \citealt{2006ApJ...651.1166M}; \citealt{2016ApJ...833...96L}).

\subsection{Geometric Radius Constraint}

The rotation period of \shorttarget\ was measured from its \emph{Spitzer}/IRAC light curves in the [3.5\,$\mu$m] and [4.3\,$\mu$m] bands by \cite{2015ApJ...799..154M}, who found an approximate period of $13 \pm 1$\,h. The probability distribution for the inclination of \shorttarget\ was explored with a Markov Chain Monte Carlo analysis based on geometry with the radius estimate and the projected rotational velocity reported in Table~\ref{tab:properties}. A non-informative prior probability on the inclination $P(i) = \sin i$ was adopted. We obtain an estimate of $i = 56_{-7}^{+11}$\textdegree. Measurements of both the photometric period and projected rotational velocity also make it possible to derive a lower limit on the radius of \shorttarget\ which ensures $v\sin i \leq v$, in a model-independent way. A 10$^7$-elements Monte Carlo approach was performed to account for the measurement errors and yielded a lower limit of $1.22 \pm 0.2$\,\rjup\ on its radius, independent of evolutionary models. Comparing this limit with the \cite{2008ApJ...689.1327S} models at the measured bolometric luminosity provides an upper limit of $150_{-130}^{+150}$\,Myr on the age of \shorttarget, independent of kinematics and consistent with membership in \replaced{AB~Doradus}{ABDMG}.

\section{CONCLUSIONS}\label{sec:conclusion}

A new trigonometric parallax based on the \emph{WISE} mission astrometry and a radial velocity measured from a GNIRS near-infrared spectrum are presented for \shorttarget\ and suggest that it is a member of the young $\sim$\,150\,Myr-old \replaced{AB~Doradus moving group}{ABDMG}. This fact alone could explain the spectroscopic peculiarities first noted by \cite{2007AJ....134.1162L} without the need to invoke an unresolved L9 + T2 spectral binary. The trigonometric parallax places \shorttarget\ in line with the sequence of other \replaced{AB~Doradus}{ABDMG} members, at a much redder $J-K$ color than field T1-T3 dwarfs and a slightly fainter $J$-band magnitude, providing more evidence that the L/T transition at a young age happens at a fainter magnitude in this particular color-magnitude space, and at a cooler effective temperature.

\acknowledgments

\added{We thank the anonymous referee of this letter for a prompt and useful report. }We thank Matthew Taylor, Sandy Leggett, Tom Geballe, and Lucas Fuhrman for their help with the GNIRS observing, and Jack Gallimore for his contributions to the DREAM-ZS fitting code. This research made use of: data products from the Two Micron All Sky Survey (2MASS), which is a joint project of the University of Massachusetts and the Infrared Processing and Analysis Center (IPAC)/California Institute of Technology (Caltech), funded by the National Aeronautics and Space Administration (NASA) and the National Science Foundation; data products from the \emph{Wide-field Infrared Survey Explorer} (\emph{WISE}), which is a joint project of the University of California, Los Angeles, and the Jet Propulsion Laboratory (JPL)/Caltech, funded by NASA. Based on observations obtained at the Gemini Observatory (science program number GN-2017B-FT-21) acquired through the Gemini Observatory Archive, which is operated by the Association of Universities for Research in Astronomy, Inc., under a cooperative agreement with the NSF on behalf of the Gemini partnership: the National Science Foundation (United States), the National Research Council (Canada), CONICYT (Chile), Ministerio de Ciencia, Tecnolog\'{i}a e Innovaci\'{o}n Productiva (Argentina), and Minist\'{e}rio da Ci\^{e}ncia, Tecnologia e Inova\c{c}\~{a}o (Brazil).

\software{BANYAN~$\Sigma$ \citep{2018arXiv180109051G}, BANYAN~II \citep{2014ApJ...783..121G}, emcee \citep{2013ascl.soft03002F}.}
\facilities{Gemini: North (GNIRS).}

\bibliographystyle{apj}

\listofchanges
\end{document}